\pdfoutput=1
\documentclass[fleqn,usenatbib]{mnras}

\usepackage{newtxtext,newtxmath}

\usepackage[T1]{fontenc}
\usepackage{ae,aecompl}

\usepackage{graphicx}	
\usepackage{amsmath}	
\usepackage{amssymb}	
\usepackage{subfigure}
\usepackage[british]{babel}             
\usepackage{upgreek}
\usepackage{gensymb}

\usepackage[pdfpagelabels=false,colorlinks]{hyperref}

\usepackage{enumitem}

\hypersetup{pdfauthor={C. A. Murray},
               pdftitle={Photometry and Performance of SPECULOOS-South},
               pdfkeywords={atmospheric effects -- techniques: photometric -- planets and satellites: detection},
               bookmarksnumbered=true}

\title[Photometry and Performance of SPECULOOS-South]{Photometry and Performance of SPECULOOS-South}

\author[C. A. Murray et al.]{C. A. Murray$^{1}$\thanks{E-mail: cam217@cam.ac.uk},
L. Delrez$^{1}$,
P. P. Pedersen$^{1}$,
D. Queloz$^{1}$,
M. Gillon$^{2}$,
\newauthor
A. Burdanov$^{2}$,
E. Ducrot$^{2}$,
L. J. Garcia$^{2}$,
F. Lienhard$^{1}$,
B. O. Demory$^{4}$,
E. Jehin$^{3}$,
\newauthor
J. McCormac$^{6,7}$,
D. Sebastian$^{2}$,
S. Sohy$^{3}$,
S. J. Thompson$^{1}$,
A. H. M. J. Triaud$^{5}$,
\newauthor
V. V. Grootel$^{3}$,
M. N. G\"unther$^{8,9}$,
C. X. Huang$^{8.9}$
\\
$^{1}$Cavendish Laboratory, JJ Thomson Avenue, Cambridge CB3 0HE, UK\\
$^{2}$Astrobiology Research Unit, Universit\'e de Li\`ege, All\'ee du 6 Ao\^ut 19C, B-4000 Li\`ege, Belgium\\
$^{3}$Space Sciences, Technologies and Astrophysics Research (STAR) Institute, Universit\'e de Li\`ege, All\'ee du 6 Ao\^ut 19C, B-4000 Li\`ege,\\ Belgium\\
$^{4}$University of Bern, Center for Space and Habitability, Sidlerstrasse 5, CH-3012 Bern, Switzerland\\
$^{5}$School of Physics \& Astronomy, University of Birmingham, Edgbaston, Birmingham B15 2TT, UK\\
$^{6}$Centre for Exoplanets and Habitability, University of Warwick, Gibbet Hill Road, Coventry CV4 7AL, UK\\
$^{7}$Dept.\ of Physics, University of Warwick, Gibbet Hill Road, Coventry CV4 7AL, UK\\
$^{8}$Department of Physics, and Kavli Institute for Astrophysics and Space Research, Massachusetts Institute of Technology, Cambridge, \\MA 02139, USA\\
$^{9}$Juan Carlos Torres Fellow\\
}

\date{Accepted XXX. Received YYY; in original form ZZZ}

\pubyear{2020}

   \volume{{\rm in press}}
\begin{document}
\label{firstpage}
\pagerange{\pageref{firstpage}--\pageref{lastpage}}
\maketitle

\begin{abstract}
SPECULOOS-South, an observatory composed of four independent 1m robotic telescopes, located at ESO Paranal, Chile, started scientific operation in January 2019. This Southern Hemisphere facility operates as part of SPECULOOS, an international network of 1m-class telescopes surveying for transiting terrestrial planets around the nearest and brightest ultra-cool dwarfs. To automatically and efficiently process the observations of SPECULOOS-South, and to deal with the specialised photometric requirements of ultra-cool dwarf targets, we present our automatic pipeline. This pipeline includes an algorithm for automated differential photometry and an extensive correction technique for the effects of telluric water vapour, using ground measurements of the precipitable water vapour. Observing very red targets in the near-infrared can result in photometric systematics in the differential lightcurves, related to the temporally-varying, wavelength-dependent opacity of the Earth's atmosphere. These systematics are sufficient to affect the daily quality of the lightcurves, the longer time-scale variability study of our targets and even mimic transit-like signals. Here we present the implementation and impact of our water vapour correction method. Using the 179 nights and 98 targets observed in the $I+z'$ filter by SPECULOOS-South since January 2019, we show the impressive photometric performance of the facility (with a median precision of $\sim$1.5\,mmag for 30-min binning of the raw, non-detrended lightcurves) and assess its detection potential. We compare simultaneous observations with SPECULOOS-South and TESS, to show that we readily achieve high-precision, space-level photometry for bright, ultra-cool dwarfs, highlighting SPECULOOS-South as the first facility of its kind.

\end{abstract}

\begin{keywords}
 atmospheric effects -- techniques: photometric -- planets and satellites: detection
\end{keywords}



\section{Introduction}

The search for extra-terrestrial life is one of the greatest challenges in modern-day astronomy, driven by the question: Are we alone in the universe? A promising path to an answer is to search for temperate Earth-sized exoplanets in order to probe their atmospheres for biosignatures with next-generation telescopes, such as the James Webb Space Telescope \citep{Gardner2006} and future Extremely Large Telescopes (e.g. \citealt{elt}, \citealt{tmt}). 

The case for ultra-cool dwarf (UCD) hosts is compelling as we move towards detecting Earth-sized, temperate worlds. UCDs are Jupiter-sized objects of spectral type M7 and later, with effective temperatures cooler than 2700\,K \citep{kirkpatrick2005}. Compared to a Sun-like host,
temperate planets around UCDs have more frequent transits, there is a higher geometric probability of observing the transit, and transit depths are 2 orders of magnitude deeper (Earth-radius planets orbiting UCDs have transit depths of $\sim$1 per cent). Due to their low luminosities and small sizes, the detection of spectroscopic signatures in the atmosphere of a temperate terrestrial planet is also more favourable for UCDs than any other host star \citep{KalteneggerTraub2009, Seager2009, deWit2013}. Despite being numerous in our local stellar neighbourhood \citep{kirkpatrick2012}, there remain many unanswered questions about UCDs, including a lack of statistics on their planet population \citep{Delrez2018}. These gaps in our knowledge, as well as the discovery of seven transiting Earth-sized exoplanets in temperate orbits of 1.5 to 19\,d around TRAPPIST-1 \citep{Gillon2016,Gillon2017} helped to strengthen the case for a survey performing dedicated photometric monitoring of UCDs, SPECULOOS (Search for Habitable Planets EClipsing ULtra-cOOl Stars, \citealt{Gillon2018, Burdanov2017, Delrez2018}), and has motivated the development of future UCD surveys (e.g. \citealt{Tamburo2019}). 

While the photometric precisions reached by ground-based transit surveys has improved dramatically over the past 20 years, these facilities are not yet able to detect the shallow 0.01 per cent transit depths produced by an Earth-radius planet orbiting a Sun-like host. Limited by the Earth's rapidly changing weather and atmospheric conditions, current state-of-the-art facilities, such as Next Generation Transit Search (NGTS, \citealt{Wheatley2018}) and SPECULOOS-South, are able to reach photometric precisions of 0.1 per cent. Ground-based transit surveys have previously shown a trade-off between two factors; the size of detectable planet and the photometric quality. While observing Sun-like objects in the visible reduces the systematics caused by the Earth's atmosphere, it limits the smallest detectable planets to Neptune-sized. On the other hand, observing redder objects, such as mid-to-late M dwarfs, which are faint in the visible and therefore must preferentially be observed in the IR or near-IR,
allows for the detection of super-Earth and Earth-sized planets. Observing these objects, however, comes with significant challenges. The stellar variability and common flares \citep{Williams2015, Gizis2017, Guenther2019} of low-mass red dwarfs can complicate the detection of transiting planets. In addition, in the near-IR the varying wavelength-dependent opacity of the Earth's atmosphere has significant effects on the incoming light. Specifically, second-order extinction effects due to highly variable absorption by atmospheric water vapour has previously limited the quality of the photometry for red dwarfs, as experienced by MEarth \citep{Berta2012}. In this paper, we present a method of modelling and correcting the effect of precipitable water vapour (PWV) during differential photometry. Not only does this correction eliminate the chance of spurious transit-like signals caused by short time-scale changes in PWV, but it significantly reduces the red noise in the photometry.

The SPECULOOS survey is a network of 1m-class robotic telescopes searching for transiting terrestrial planets around the nearest and brightest UCDs. The main facility of the network in the Southern Hemisphere, the SPECULOOS-South Observatory (SSO), started full scientific operations in January 2019 at the ESO Paranal Observatory (Chile). The SPECULOOS-North Observatory (SNO) based at the Teide Observatory in Tenerife (Canary Islands) is currently being developed and saw the first light of its first telescope in June 2019. Along with contributions from SAINT-EX in San Pedro M\'{a}rtir (Mexico), TRAPPIST-South at ESO's La Silla Observatory (Chile), and TRAPPIST-North at the Ouka\"{i}meden Observatory (Morocco) \citep{Gillon2011,Jehin2011}, these observatories will work together to observe around approximately 1200 of the nearest and brightest UCDs. Over the course of the next ten years, this survey will allow us to determine the frequency and diversity of temperate terrestrial planets around low-mass objects and will provide first-class targets for further atmospheric study in the search for signs of habitability beyond the Solar System.

The SSO aims to detect single transits from Earth-sized planets, requiring photometric precisions of $\sim$0.1 per cent. To obtain the necessary high signal-to-noise ratio (SNR) lightcurves, and to deal with the specificity of our very red targets, we developed a specialised automatic pipeline to process and reduce the data from the SSO. This pipeline includes a novel differential photometry algorithm and a correction of the effects of variable telluric water absorption. Since the start of scientific operations, we have been tracking the quality of the SSO's photometry. This provides feedback into the photometric pipeline and allows us to assess whether the facility is reaching the expected performances set out by the survey goals. This paper details the various stages involved in assessing SSO's performance during its first year of operation: a description of the SPECULOOS-South Pipeline in Section \ref{Pipeline}, the differential photometry technique developed in Section \ref{diff_photom}, the impact of telluric water vapour on photometry, and an outline of the implemented correction, in Section \ref{pwv}, and the determination of the overall photometric performance of the survey in Section \ref{performance}.

\section{The SPECULOOS-Southern Observatory}

The SSO consists of four robotic 1-m Ritchey--Chretien telescopes\footnote{The SSO telescopes are named after the four Galilean moons: Europa, Io, Callisto and Ganymede. This is partially because this Jovian system mirrors the size ratio between Earth-sized planets and their UCD host, but also as a tribute to the first objects discovered to orbit a body other than the Earth, challenging the geocentric Ptolemaic model of the time.}, each equipped with a deeply depleted CCD detector which is optimised for the near-IR. For the vast majority of our observations we use the $I+z'$ custom-designed filter (transmittance >90 per cent from 750\,nm to beyond 1000\,nm) due to the faintness of our red targets in the optical wavelength domain. However, we are limited beyond 950\,nm by the quantum efficiency of our CCD detector. Further technical information is shown in Table \ref{tab:SSO} and described in more detail, alongside transmission curves (Figure 7 for the $I+z'$ filter transmission curve and Figure 6 (right) for the total efficiency in $I+z'$), in \cite{Delrez2018}.

\begin{table}
 \caption{Technical specifications of each telescope in the SSO}
 \label{tab:SSO}
 \begin{tabular}{|p{2cm}|p{5.55cm}}
  \hline
  & Specification\\
  \hline
  Mirrors & 1\,m diameter primary with a $f/2.3$ focal ratio and 28\,cm diameter secondary. Combined $f/8$ focal ratio. Both mirrors are coated with pure aluminium.\\
  Camera & Andor iKon-L thermoelectrically-cooled camera \\
  CCD Detector & Near-IR-optimized deeply-depleted  2k\,$\times$\,2k  e2v  CCD  detector \\
  CCD Quantum Efficiency & $\sim$350 (near-UV) to $\sim$950\,nm (near-IR). Peak quantum efficiency of 94\% at 740\,nm.\\
  Field of View & 12\,$\times$\,12\,arcmin$^2$ \\
  Pixel Scale & 0.35\,arcsec\,pixel$^{-1}$ \\
  Pixel Size & 13.5\,\micro m \\
  Dark Current & $\sim$0.1 e$^-$\,s$^{-1}$\,pixel$^{-1}$ when the camera is operated at \hbox{$-60\degree$C.} \\
  Readout Mode & Usually 1MHz readout mode with a pre-amplifier gain of 2\,e$^-$\,ADU$^{-1}$ providing readout noise of \hbox{6.2\,e$^-$} \\
  Gain & 1.04\,e$^-$\,ADU$^{-1}$\\
  Filter Wheel & Finger Lakes Instrumentation (model CFW3-10) allowing ten 5 x 5\,cm filters.\\
  Filters & All telescopes: Sloan-$g'$, -$r'$, -$i'$, -$z'$, $I+z'$ ,`blue-blocking' filters. Selected telescopes: broad-band Johnson--Cousins $B$, $R_{C}$ and $V$ filters, and the Sloan $u'$ filter.\\
  \hline
 \end{tabular}
\end{table}

Observations on the four telescopes are started remotely each night. Each telescope operates independently and in robotic mode following plans written by SPECULOOS's automatic scheduler. On average 1--2 targets are observed by each telescope per night. Each target will be observed continuously for between several hours and an entire night (for however long weather permits and the target is observable). Typically we observe each target between 1 and 2 weeks, depending on its spectral type, so as to efficiently probe the temperate region around that object. As this is a targeted survey, our targets are spread over the sky, therefore there is only one target per field of view. During operation, each telescope uses the auto-guiding software, \textsc{donuts} \citep{McCormac2013}, to calculate real time guiding corrections and to re-centre the telescope pointing between exposures. Systematic errors caused by the drift of stars on the CCD (with inhomogeneous pixel response) can severely limit the precision of time-series photometry, therefore fixing stellar positions at the sub-pixel level is essential. \textsc{donuts} is also capable of auto-guiding on defocused stars, useful, for example, when we observe bright objects. 

All raw images recorded by the facility are automatically uploaded at the end of the night to the online ESO archive\footnote{\url{http://archive.eso.org/eso/eso_archive_main.html}}. These images are then automatically downloaded to a server at the University of Cambridge (UK), and analysed by the pipeline. All images (and extracted lightcurves of all objects observed in all fields) will be made publicly available after a 1-year proprietary period.

\section{The SPECULOOS-South Pipeline} \label{Pipeline}

Every survey presents unique calibration and photometric challenges and so we have developed a pipeline specific for SSO. We designed this photometric pipeline to be fast, automatic, and modular. Depending on the targets and conditions of the night, we accumulate approximately between 250 and 1000 images per telescope per night with typical exposure times of 10--60s, corresponding to between 4 and 16 GB of data. Flexibility in the pipeline allows us to perform various quality checks, extract feedback and use these to optimise the performance of the survey. Modularity allows reprocessing certain stages of the pipeline with improved algorithms, without requiring a full rerun.

The structure and data format of the SSO pipeline is based on the architecture of the NGTS pipeline described in \cite{Wheatley2018}. Similarly to NGTS, we built our pipeline on top of the \textsc{casutools}\footnote{\url{http://casu.ast.cam.ac.uk/surveys-projects/software-release}} package of processing tools for image analysis, astrometric fitting and photometry \citep{Irwin2004}.

\begin{figure}
 \includegraphics[width=\columnwidth]{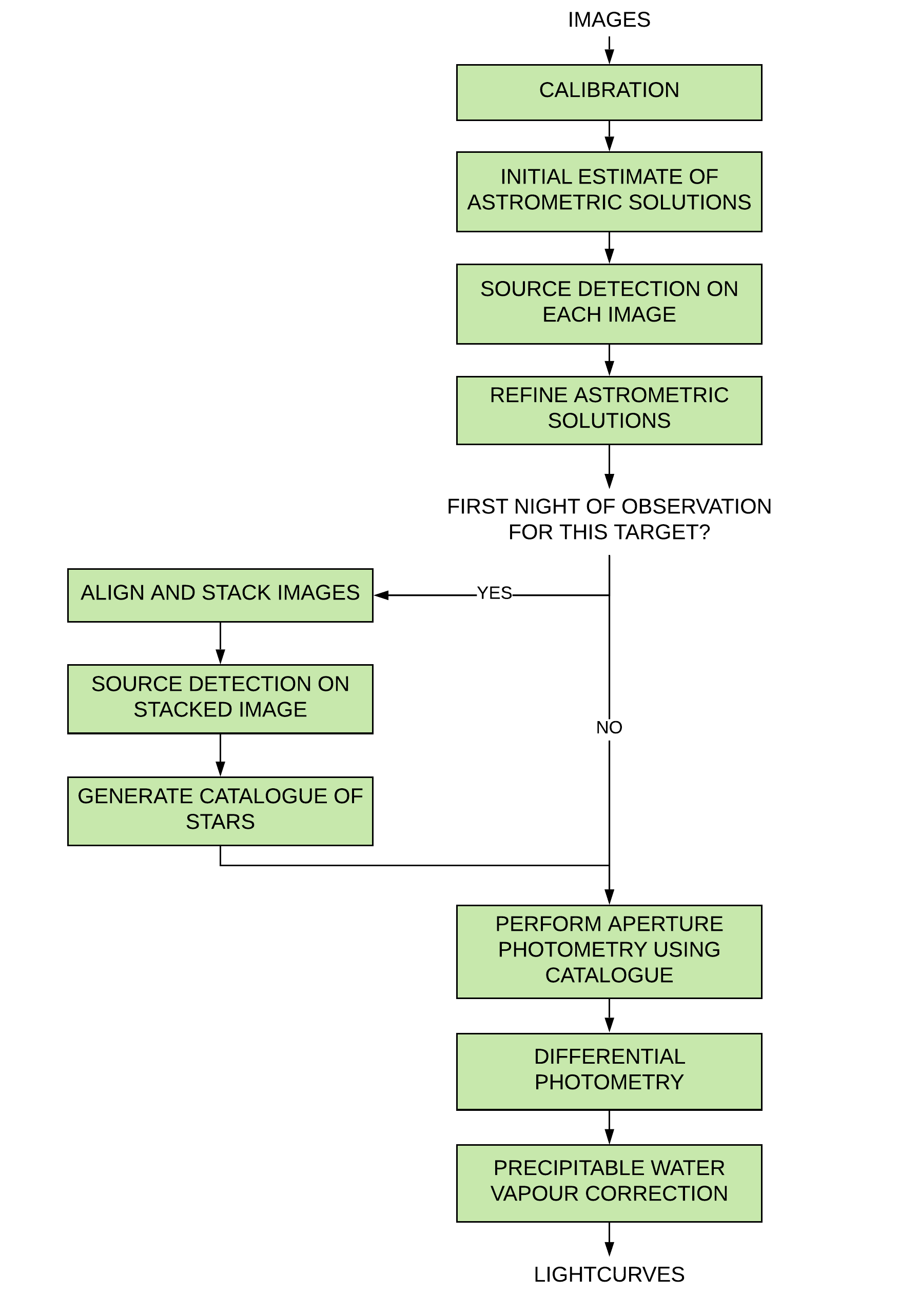}
 \caption{Simplified flowchart of the SPECULOOS-South Pipeline.}
 \label{fig:flow}
\end{figure}

The various steps of the pipeline are illustrated in Fig. \ref{fig:flow}. The science images are calibrated through bias and dark subtraction and flat-field division (Section \ref{reduction}). Astrometric solutions are found for each image (Section \ref{astrometry}). If this is the first night of observation for a given field then these images are aligned and stacked to create a stacked image. Sources detected on this stacked image are used to generate a catalogue of stars for this field of view (Section \ref{catalogue}). Precise aperture photometry measurements are extracted from each image using the catalogue with a selection of different aperture sizes (Section \ref{ap_photom}). We can then generate differential lightcurves for any object in the catalogue; either for a single night or over many nights to assemble a `global' lightcurve (Section \ref{diff_photom}). Global lightcurves can be used to assess the photometric variability of a target over multiple nights. Systematic effects, such as those caused by changes in PWV, are then removed (Section \ref{pwv}). 

\subsection{Data reduction} \label{reduction}
Standard methods of bias and dark subtraction and flat-field correction are used to calibrate the science images. Bias and dark images are taken at dawn, after the closure of the telescope dome, whereas the flat-field images are captured using the twilight sky at both dusk and dawn. All images are overscan subtracted and the bias, dark and flat-field frames are sigma-clipped and median-combined to produce master calibration images, with appropriate corrections using the other master calibration images. The master flat images are monitored over time to assess their quality and flag significant variations (e.g. moving dust). These master calibration images are then used to calibrate the science images. 

\subsection{Astrometry} \label{astrometry}
Despite good performance of the telescope guiding with DONUTS, there remain very small drifts in object positions during the night, of the order of $\sim$0.1\,arcsec ($\sim$0.3\,pixels). Precise astrometric solutions are needed for each image to accurately place apertures for photometric measurements. A local version of \textsc{astrometry.net} code \citep{Lang2010} is used to cross match each science image with reference catalogues built from the 2MASS catalogue to find an initial approximate World Coordinate System (WCS) solution. This solution is then refined by using first \textsc{imcore}, to detect sources on the image, and then \textsc{wcsfit}, to produce the final WCS solution, from the \textsc{casutools} package.

\textsc{imcore} performs source detection on an image by first computing a low resolution background image. This is done by estimating background values for 64\,$\times$\,64 pixel$^2$ sections by using an algorithm based on a robust (MAD) iterative $k$-sigma clipped median. These background values are then filtered to produce the low resolution background image. Using bi-linear interpolation, the local sky background of every pixel in the original image can then be derived. To identify a source, the algorithm searches for a connected series of 6 pixels with values higher than a user-specified threshold above the background. For the purpose of astrometry we want to use as many stars as possible, therefore we use a low limit of 2-sigma above the background sky level to detect sources.

\textsc{wcsfit} uses the initial WCS solution to further correct each image's WCS solutions for translations, skews, scales and rotations by crossmatching the sources from \textsc{imcore} with the Gaia Data Release 1 Catalogue \citep{Gaia2016}.

\subsection{Catalogue generation} \label{catalogue}
For each field of view that is observed (i.e. each target), the pipeline requires an input catalogue with the RA and DEC of the stars on which to extract aperture photometry data for each image. This catalogue is generated from a stacked image produced from 50 images in the middle of the night (in order to reduce the airmass and sky background), taken on a target's first night of observation. We have a unique catalogue for each field of view which is then referenced across all the subsequent nights that target is observed in order to track these stars over long periods of time. This catalogue is cross-matched with Gaia Data Release 2 \citep{Gaia2018} to apply proper motion corrections on a night-by-night basis. There is also the facility to cross-match with other catalogues, such as 2MASS \citep{Skrutskie2006}.

The \textsc{imstack} and \textsc{imcore} programs from the \textsc{casutools} package \citep{Irwin2004} are used in generating this catalogue. For each of the 50 science images \textsc{imstack} aligns (using the WCS solutions from \textsc{wcsfit}) and stacks these images to produce the final stacked image. 

\textsc{imstack} defines a WCS reference grid using the first image and subsequent images are then aligned and resampled on to this grid. The sigma-clipped mean of the pixel values from all images, scaled by their exposure times, is computed and recorded in the output stacked image. Outliers (defined by threshold values of 5 sigma) are removed from the averaging. \textsc{imstack} uses a bi-linear interpolation approach where an input pixel is divided into the four pixels on the output grid that surround the input equatorial position, as this can reduce systematic errors \citep{Mighell1999}. The fraction in each output pixel corresponds to the amount of overlap of the input pixel. The final stacked images are crucial in the creation of the catalogues that define each field of view. Therefore quality checks implemented by the automatic pipeline help to ensure the stacked image is created on a night with good seeing and atmospheric conditions, and ideally no defocusing, to increase the accuracy of the source positions on the field.

 \textsc{imcore} then performs source detection on the stacked image to create a catalogue of the stars in the field of view. This time, however, \textsc{imcore} searches for sources with more than 6 contiguous pixels containing counts 8-sigma above the background sky level. This higher threshold limits the detected objects to $I+z'$-magnitudes brighter than $\sim$21. The background sky level present in the stacked image will vary depending on the angular proximity and phase of the moon, however, we don't see any noticeable variation in the number of stars in the catalogue corresponding to the moon cycle, potentially due to the small pixel size of our CCDs.

\subsection{Aperture photometry} \label{ap_photom}
\textsc{imcorelist}, a fourth \textsc{casutools} program, is used to perform aperture photometry on each science image. It carries out essentially the same process as \textsc{imcore} but requires an input list of equatorial positions, provided by the catalogue, to define the positions of the apertures. \textsc{imcorelist} takes photometric measurements of each source on every image for 13 apertures sizes which are multitudes of the user-defined radius \textit{rcore} (default 4 pixels or 1.4\,arcsec)\footnote{The 13 apertures used are multiples (1/2, 1/$\sqrt{2}$, 1, $\sqrt{2}$, 2, 2$\sqrt{2}$, 4, 5, 6, 7, 8, 10 and 12) of \textit{rcore}.}. The final aperture for a given night is chosen to balance minimizing the `average spread' and correlated noise in the target's final differential lightcurve. The `average spread' of the target's differential lightcurve is defined to be the average standard deviation inside 5-min bins. We chose to minimize the RMS inside the bins multiplied by the RMS of the binned lightcurve to avoid minimising genuine photometric structure in the lightcurve (e.g. stellar variability), whilst also avoiding adding correlated noise in the lightcurve, for example from the changing FWHM and airmass during the night if we choose an aperture that is too small.

\section{Differential photometry} \label{diff_photom}

\begin{figure*}
    \centering
    \subfigure[Clear night, 2017 October 6. ]{\includegraphics[width=.48\textwidth]{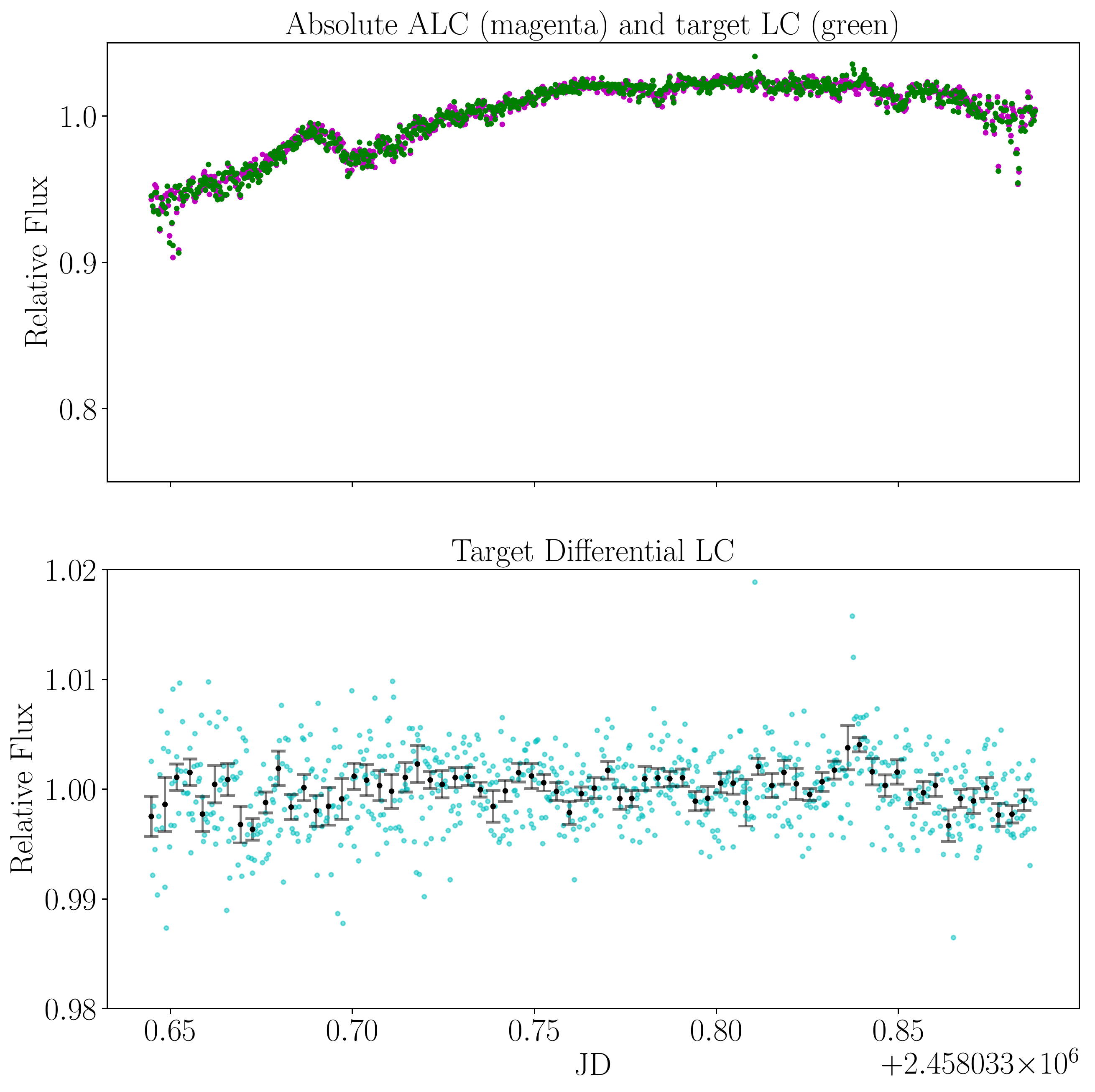}}
    \subfigure[Cloudy night, 2017 October 8.]{\includegraphics[width=.48\textwidth]{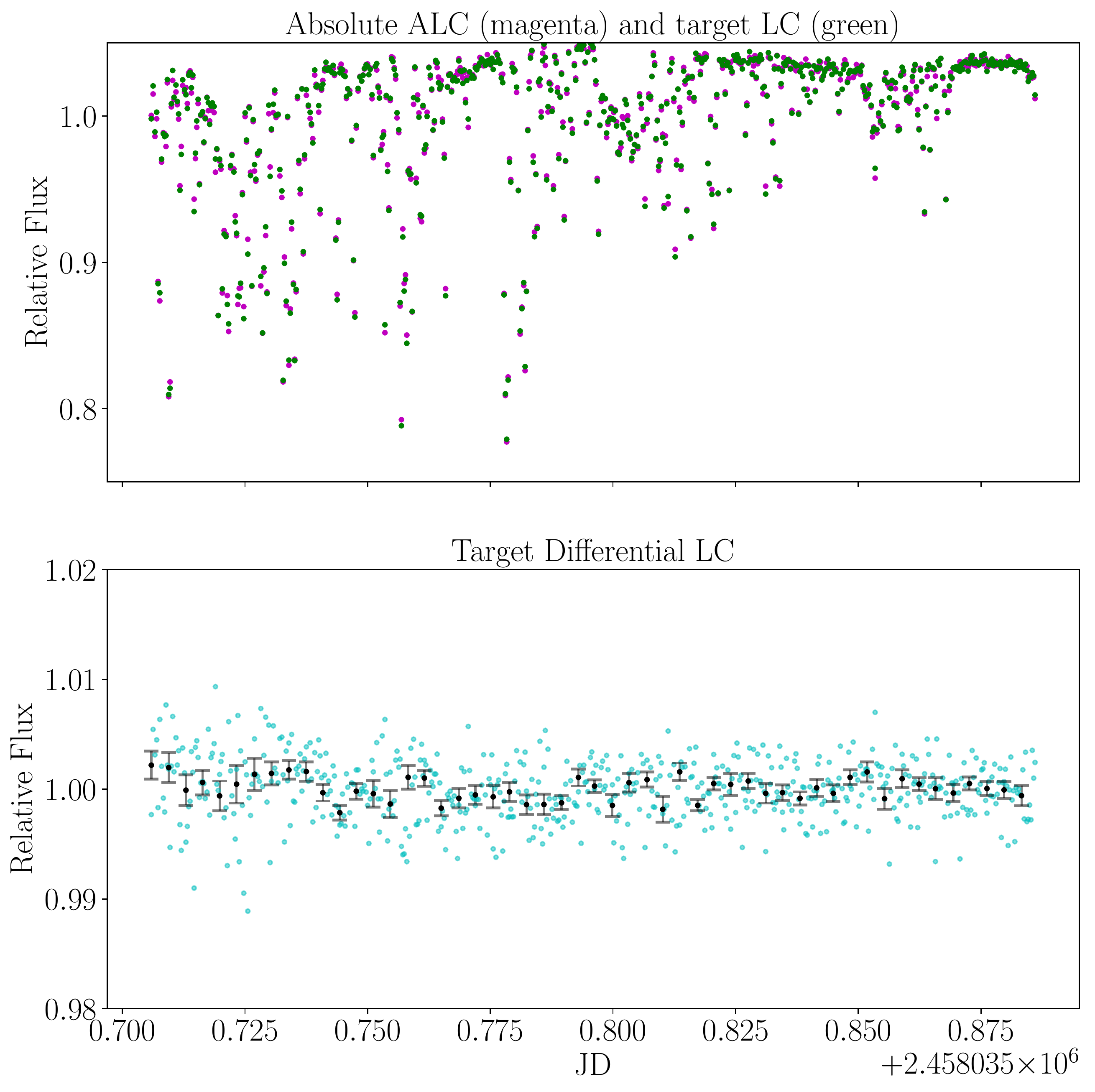}}
    \caption{Demonstration of the differential photometry algorithm on a bright M8V ($J=10.4$\,mag) target star, observed by Europa during its commissioning phase, comparing the results on a relatively clear night (a), and a cloudy night (b). The top plots show the ALC (magenta) compared to the target's absolute lightcurve (green), for both nights the optimal aperture is 11.3 pixels. The bottom plots show the target's final differential lightcurve (unbinned points in cyan and 5-min binned points in black), produced by dividing the target's absolute lightcurve by the ALC. The differential lightcurve for (a) shows a small flare-like structure (JD 2458033.84), which would be difficult to extract from the absolute lightcurve.}
    \label{fig:alc}
\end{figure*}

Differential photometry is a technique based on the assumption that stars of similar brightness and colour in a field of view will experience a common photometric pattern, due to shared atmospheric and instrumental effects. For the SSO, we developed an algorithm to automatically choose and combine multiple comparison stars to ensure that the final differential lightcurves would be reproducible and to avoid the time-intensive, manual selection of stars and potential observer bias. Statistically, it is optimal to use as many stars as possible, weighted appropriately, to reduce the noise levels in the final differential lightcurves. The algorithm implemented in our  pipeline is based on a concept described in \cite{Broeg2005}. This iterative algorithm automatically calculates an `artificial' comparison lightcurve (ALC) by weighting all the comparison stars accounting for their variability, and removing those which are clearly variable. To optimise our pipeline for SSO data, several major changes from the algorithm developed by \cite{Broeg2005} were implemented. The basic algorithm is described at the beginning of Section \ref{ALC}, while our implemented changes are described in Sections \ref{varcut} to \ref{faint}. A demonstration of the need for differential photometry and the correction with the ALC on observation nights of different quality is shown in Fig. \ref{fig:alc}.

\subsection{Generating an `artificial' comparison star} \label{ALC}
The following method is similar to that described in \cite{Broeg2005} where each object (excluding the target and any saturated stars), $i$, is assigned a weight, $W_{\textrm{var},i}$, determined by its variability.

\begin{enumerate}[leftmargin=.5cm]
    \item 
    The initial weights are defined as:
    \begin{equation}
        W_{\textrm{var},i} = 1/ \sigma_{\textrm{photon},i} ^2
    \end{equation}
     where $\sigma_{\textrm{photon},i}$ is the photon noise of star $i$, therefore in this step $W_{\textrm{var},i}$ is set to be equal to the average flux for each object. These weights are normalised such that they sum to 1. \\
    \item The ALC is constructed from the weighted mean of the normalised flux ($F$) of each of the $n$ objects in the field, at each frame $j$:
        \begin{equation} \label{eq:ALC}
            ALC_j = \frac{\sum_{i=1}^{n}W_{\textrm{var},i} F_{ij}}{\sum_{i=1}^{n}W_{\textrm{var},i}}
        \end{equation} 
    \item Every star's absolute lightcurve, $F$, is divided by this ALC to produce a differential lightcurve. \\
    \item The weight for star $i$ is replaced by: 
        \begin{equation} \label{eq:W}
            W_{\textrm{var},i} = 1/ \sigma_i ^2
        \end{equation}
    where $\sigma_i$ is the standard deviation of the differential lightcurve for star $i$.
\end{enumerate}
Stages (ii), (iii) and (iv) are repeated with these new weights until the weights are constant to within a threshold of 0.00001.

\subsubsection{Initial variability cut}\label{varcut}

From testing it became clear that if there was variability in the brightest stars, which are highly weighted during stage (i) of this algorithm, then the initial ALC estimate would be significantly affected. If these objects  are not removed, in the next iteration, they would weight down stable stars and weight up those with any similar time variability structure. This results in a runaway effect, down-weighting the more stable comparison stars. Therefore we simply included a variability check prior to generation of the initial ALC by sigma-clipping across all stars' normalised lightcurves for each frame. If any object has >20 per cent of its values clipped it is determined that this object is variable, and it is removed.

\subsubsection{Colour}\label{colour}
\begin{figure}
 \includegraphics[width=\columnwidth]{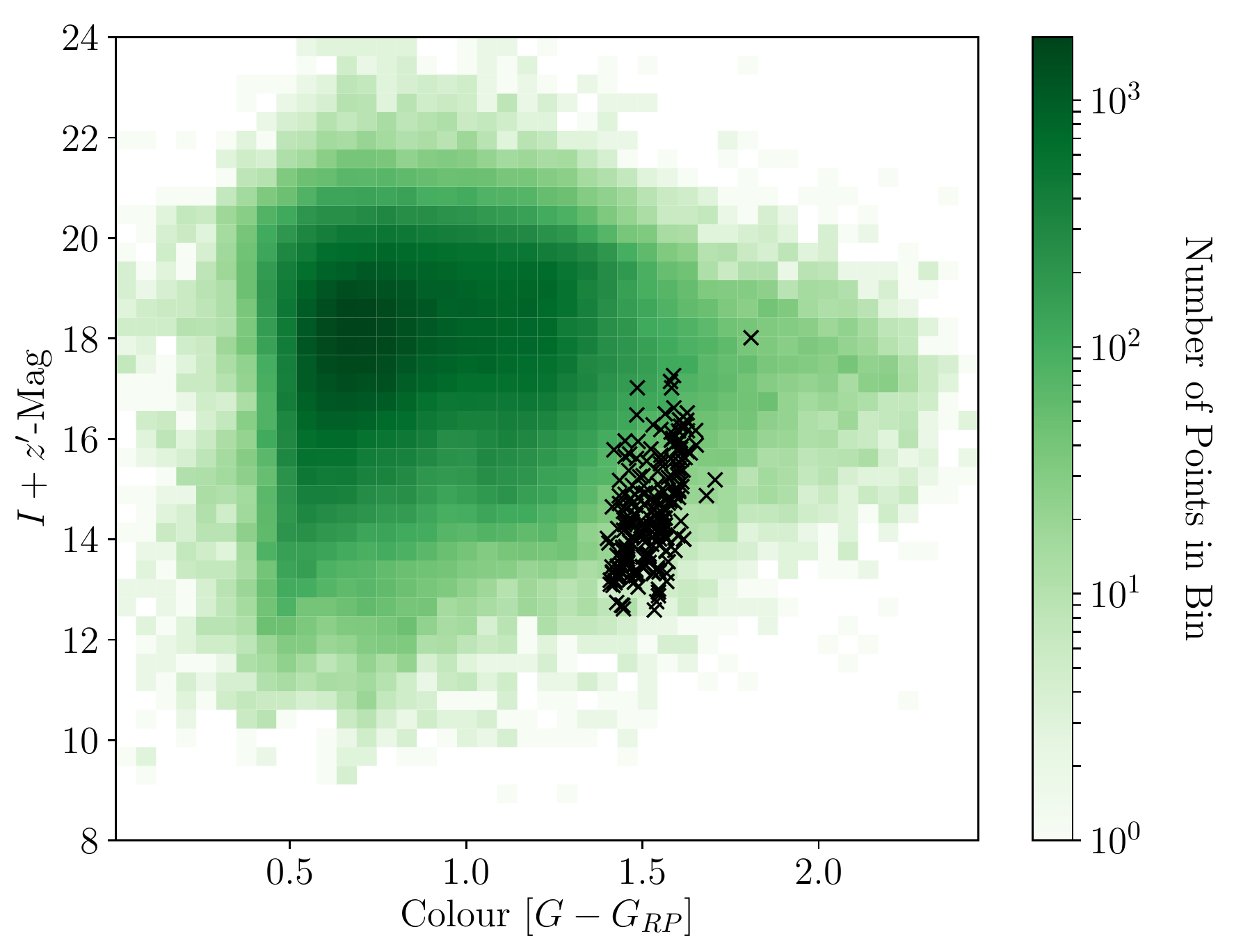}
 \caption{$I+z'$-Magnitude against Gaia colour, $G-G_{\textrm{RP}}$, for all catalogued stars in every observed field of view (on all telescopes) since 2017 April. The SSO targets are marked by black crosses}
 \label{fig:colmag}
\end{figure}

By design, the SSO's targets are usually among the reddest stars in the field of view (FOV), and so there is always a colour mismatch between the target star and the comparison stars (see Fig. \ref{fig:colmag}), resulting in second-order differential extinction effects. The redder comparison stars in the field are often significantly dimmer than the target. We therefore resisted the temptation to implement a strict cut of the bluest (and brightest) stars, which would increase the noise in the ALC, and subsequently the target's differential lightcurve. Instead we decided to correct the differential extinction in a later stage of the pipeline (see Section \ref{pwv}).

\subsubsection{Distance}\label{dist}
Due to spatially varying atmospheric and optical effects, we added an additional weight based on projected distance from the target star, using the formula:
\begin{equation}
    W_{\textrm{dist}, i} = \frac{1}{1 + \left(\displaystyle\frac{a s_{i}}{s_{\textrm{max}}}\right)^2}
\end{equation}
where $W_{\textrm{dist}, i}$ is the distance weight of star $i$, $s_{i}$ is its separation from the target star, $s_{\textrm{max}}$ is the maximum distance of any star from the target and $a$ is a parameter optimised for each night. We chose this form to be finite and relatively flat near the target object and decay slowly as the distance on sky increases.
The value of $a$ is chosen to minimize the `average spread' of the target's differential lightcurve (as defined in Section \ref{ap_photom}). 
We normalise these weights to sum to 1, and combine the distance weights and the variability weights from Section \ref{ALC}, $W_{\textrm{var, i}}$, to produce the final weights used in the ALC:
\begin{equation}
    W_{i} = W_{\textrm{var},i}W_{\textrm{dist}, i}
\end{equation}
Once again, we normalise these weights, which then replace the weights in step (iv) of the iteration process.

\subsubsection{Removal of the faintest stars}\label{faint}
Ideally, we would use as many comparison stars as possible (weighted appropriately), however, we found that including a large number of faint comparison stars tends to increase the noise in the ALC. It is particularly clear on nights where the atmospheric transmission varies by more than 30 per cent, suggesting passing clouds or poor weather conditions which limit our ability to conduct precise photometric measurements. It was therefore necessary to include a threshold that could be adjusted each night, to remove a certain number of faint stars. This threshold value is chosen automatically to minimize the `average spread' of the target's final differential lightcurve (as defined in Section \ref{ap_photom}).

\subsection{Night and global lightcurves}
 
Rather than treating every night of data independently, we can perform the previous differential photometry process (see Section \ref{ALC}) on longer duration photometric time-series. This allows us to study photometric variability and rotation over periods of time longer than a night.

To create the global lightcurves, we apply the differential photometry algorithm to the entire time series at once, which can span several nights, weeks or months. To ensure any observed changes in flux between nights are caused by real astrophysical variability (and not as a consequence of the differential photometry process) we use the same comparison stars, weightings and aperture across all nights. This decision, however, reduces our ability to optimise per night, which may result in residuals in the target's final differential lightcurve, which are particularly obvious on nights with sub-optimal observing conditions.

Choosing the optimal aperture for the global lightcurves is not a straightforward  process. The optimal aperture changes from night to night, mostly due to seeing variations affecting the full width at half maximum (FWHM) of the point spread function (PSF) of sources on the field of view. In practice, the optimal aperture of the series has to be large enough to avoid loosing stellar flux on the nights with larger seeing. This, however, tends to increase the background noise, which disproportionately affects the faintest stars. This effect is mitigated by the cut we implemented on the faintest stars (see Section \ref{faint}).

\subsection{Bad weather flag}

\begin{figure}
 \includegraphics[width=\columnwidth]{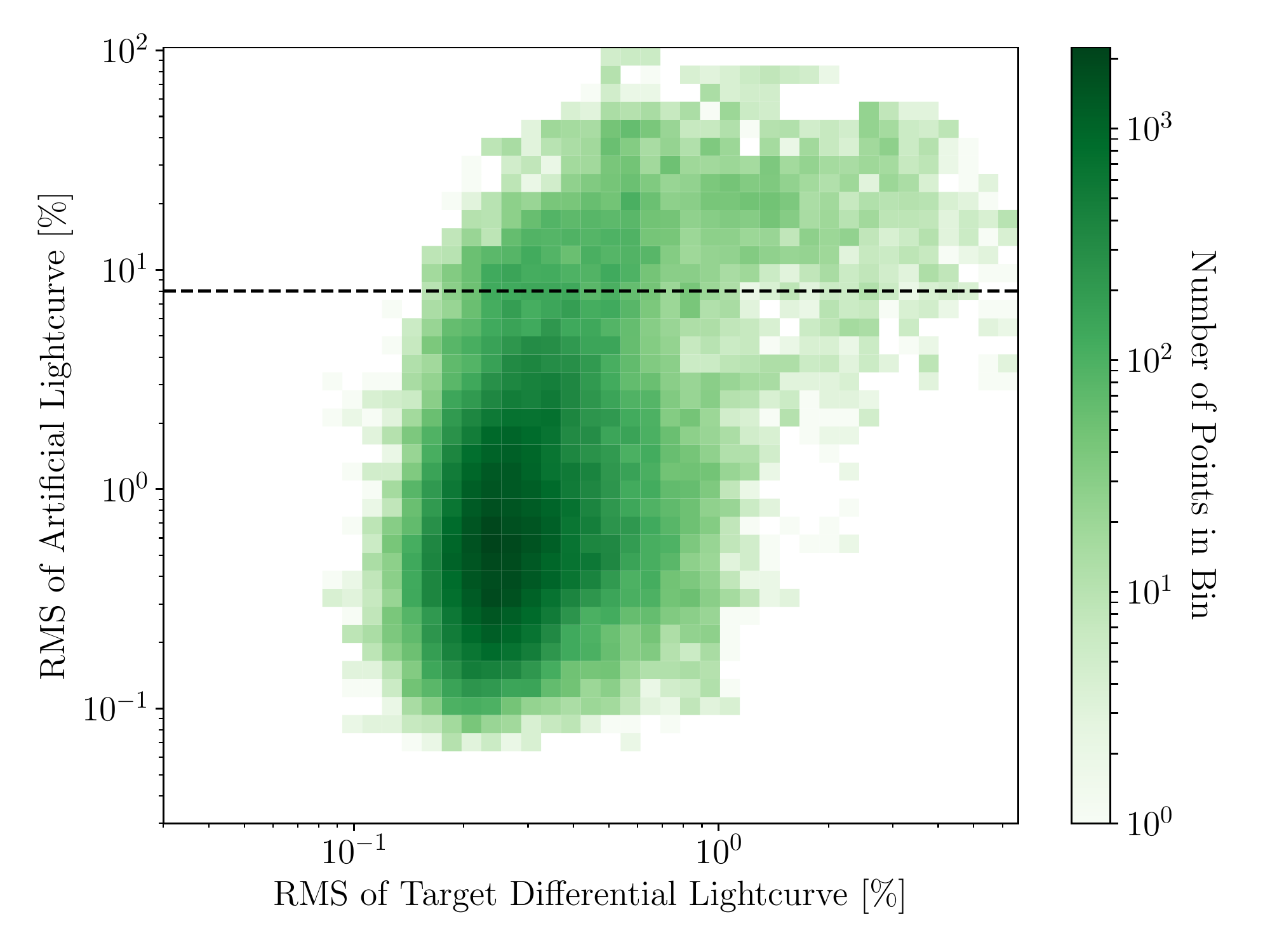}
 \caption{Local RMS of unbinned artificial lightcurves (ALC) against local RMS of unbinned target differential lightcurves for all of Io's observations since 2019 January 1. For this analysis, there is no water vapour correction or removal of cosmic hits, flares or variability, which may cause a points at low ALC RMS but a high target lightcurve RMS of a few per cent. It is clear there is an much larger variation in quality of the target differential lightcurves when the local RMS of the ALCs exceeds the threshold of $\sim$8 per cent, shown by the black dashed line.}
 \label{fig:rms_thresh}
\end{figure}

``Bad weather'' in the context of the pipeline is defined as the point at which the observing conditions of the night have a significant impact on the target's differential lightcurve. It is not related to any specific external monitoring of the weather. While in theory the ALC should allow us to correct for any change of atmospheric transmission, empirically there is a practical limit to this assumption. We found there was a threshold for the local RMS of a data-point in the ALC, above which the local RMS of the corresponding data-point in the target's differential lightcurve increased dramatically. The local RMS of a given data-point in the lightcurve is defined as the RMS measured when considering a time range (or box) of $\pm$0.005\,d ($\sim$7.2\,min) around that point in time. Combining many nights of data allowed us to determine a threshold of 8 per cent to flag (not remove) bad weather in the lightcurves (see Fig. \ref{fig:rms_thresh}).

\section{Telluric Water Vapour} \label{pwv}

\begin{figure}
 \includegraphics[width=\columnwidth]{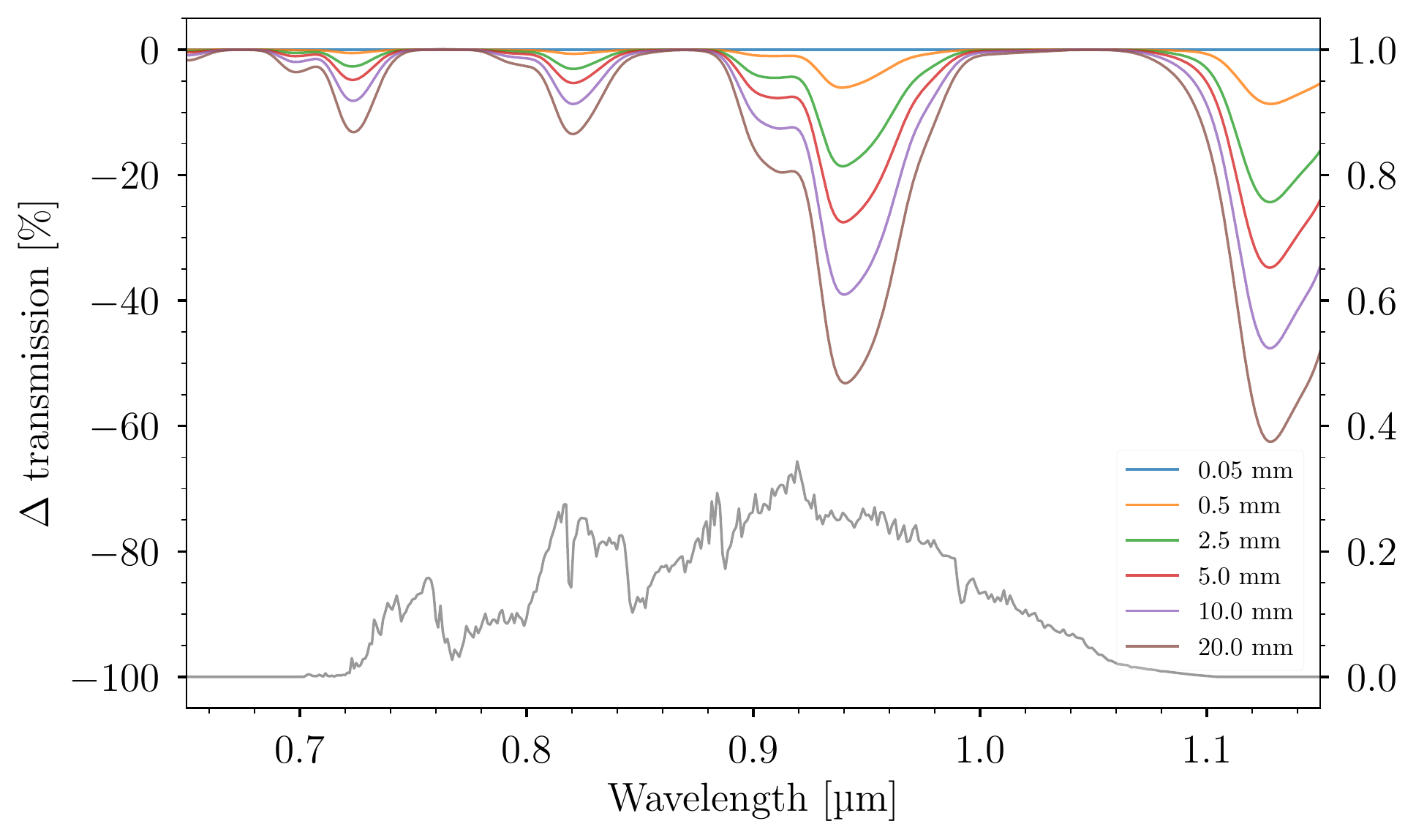}
 \caption{Percentage change in atmospheric transmission (left-hand axis) at different increases of PWV (from 0.05\,mm) and, in grey, the spectrum of TRAPPIST-1 as observed through the SSO's $I+z'$ filter, taking into account the overall system efficiency (right-hand axis), as described in \protect\cite{Delrez2018}.}
 \label{fig:transmission}
\end{figure}

\begin{figure}
 \includegraphics[width=\columnwidth]{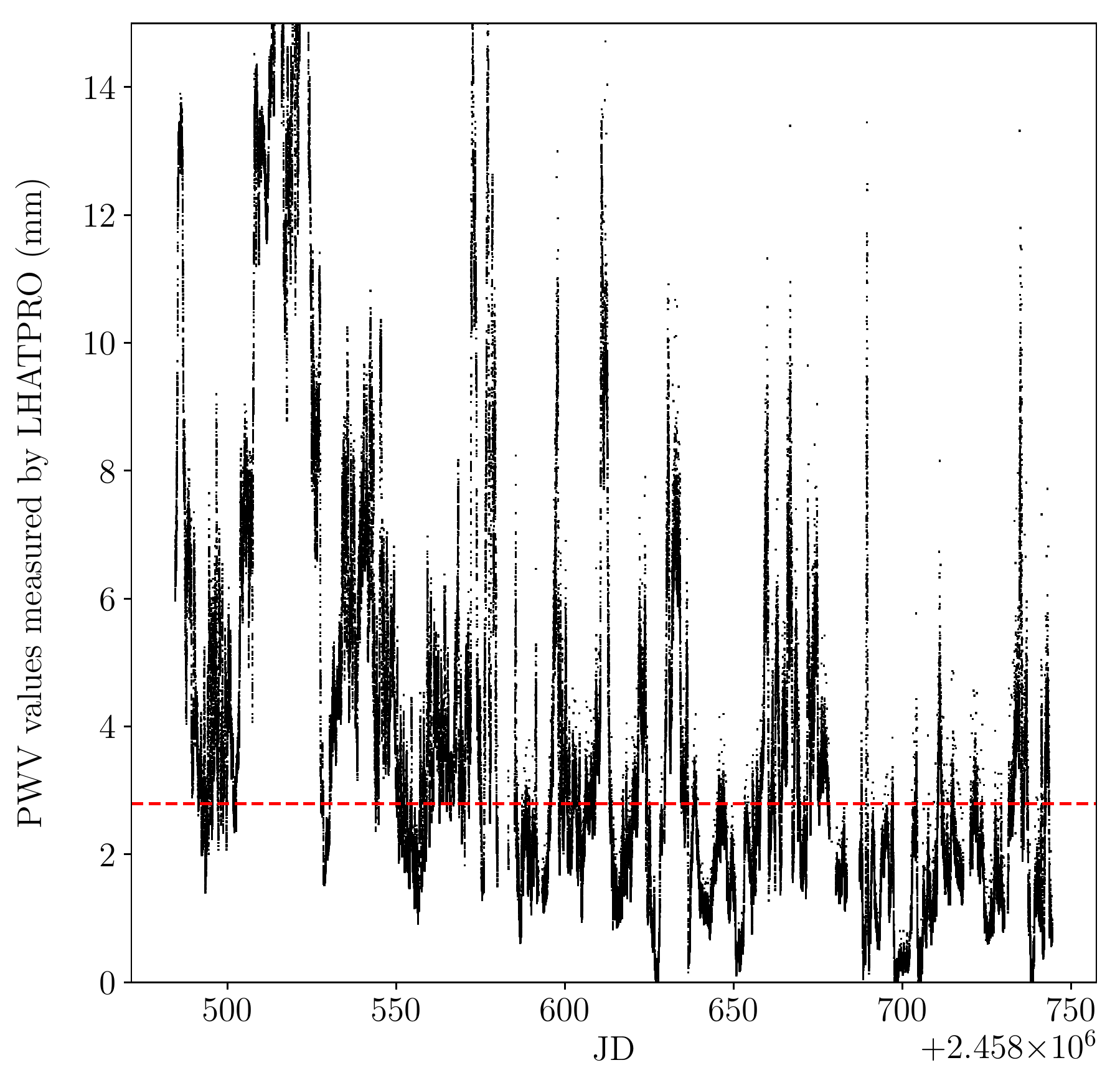}
 \caption{PWV variations in Paranal, measured by LHATPRO from 2019 January 1 to 2019 September 18. The median value of 2.795\,mm is shown by the dashed red line.}
 \label{fig:PWV_variations}
\end{figure}

SPECULOOS faces additional photometric challenges to most other ground-based transit surveys, as we are observing very red objects in the near-IR. For the vast majority of our observations, we use the $I+z'$ photometric filter. This wavelength range is strongly affected by atmospheric water absorption lines, and to a much lesser extent by OH radical absorption and emission (airglow) lines. The atmospheric transmission varies strongly with the amount of precipitable water vapour in the Earth's atmosphere (see Fig. \ref{fig:transmission}), which can be measured from the ground. Despite the fact that Paranal is an exceptionally dry site (Chilean Atacama Desert), with a nightly median PWV of $\sim$2.4\,mm and 45 nights a year less than 1\,mm of PWV \citep{Kerber2014}, it can experience large variations in PWV. This includes pronounced seasonal variations \citep{Kerber2010}, and variations of up to 20\,mm over long time-scales and even as much as 13\,mm during a single night of observation (see Fig. \ref{fig:PWV_variations}).

\begin{figure}
 \includegraphics[width=\columnwidth]{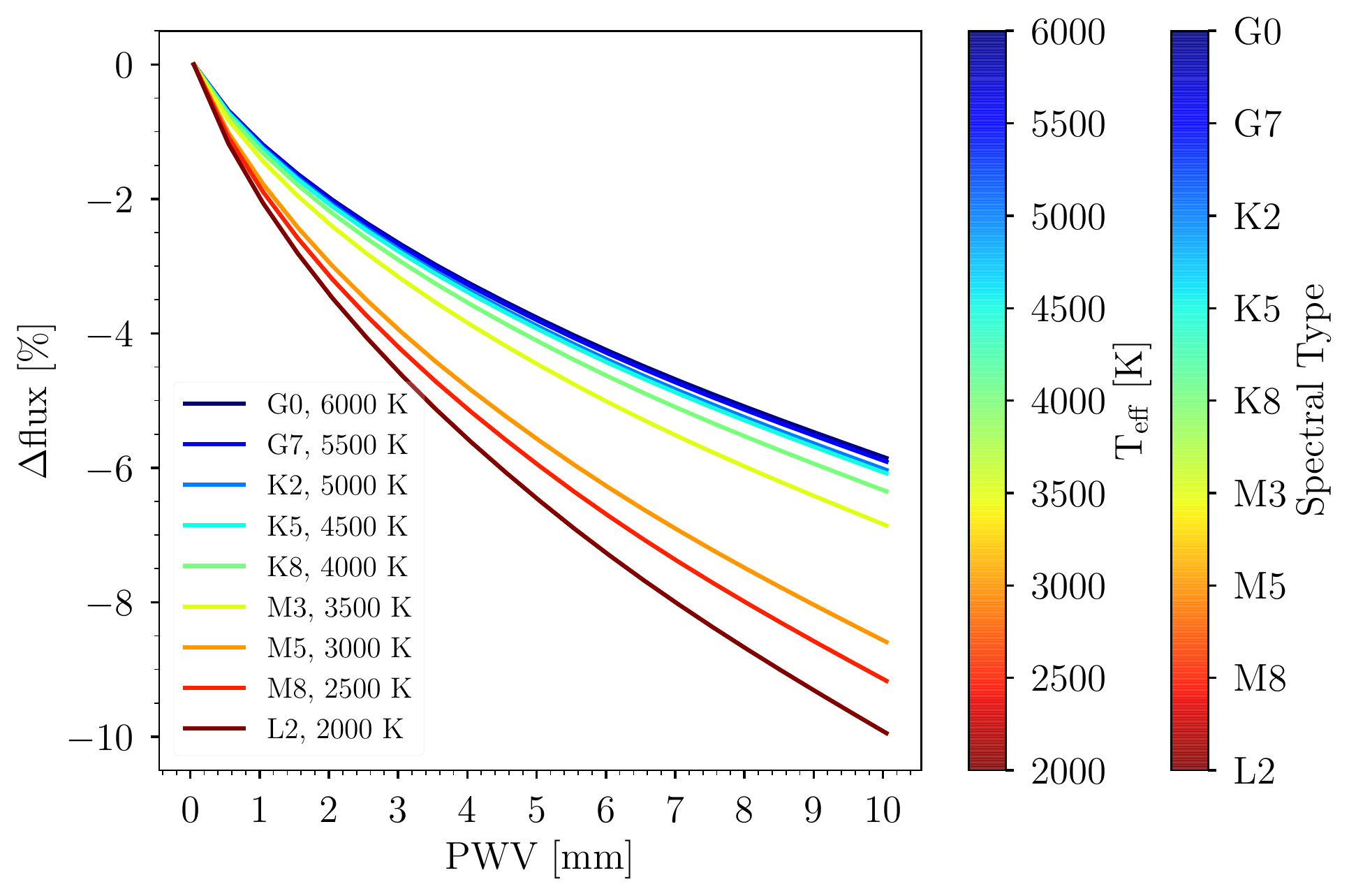} \caption{Demonstration of the differential flux effect in the $I+z'$ band with changing PWV. For example, an M8 target star will experience a 9 per cent flux drop for a PWV change from 0.05 to 10\,mm, whereas G- and K-type comparison stars (the difference is minimal between hotter stars) will only experience a 5--6 per cent flux decrease.}
 \label{fig:pwv_spectype}
\end{figure}

By construction of the SPECULOOS's UCD survey, there is always some mismatch in spectral type (and thus colour) between the target and comparison stars used to perform  differential photometry. Since redder wavelengths are more readily absorbed by water than bluer wavelengths, when the amount of PWV in the atmosphere changes then objects of different spectral types (whose spectral energy distributions peak at different wavelengths) will experience differing amounts of atmospheric absorption (see Fig. \ref{fig:pwv_spectype}). Temporal variations in PWV can therefore imprint second-order extinction residuals on the target differential lightcurves during differential photometry of order $\sim$1 per cent \citep{Baker2017} or more, when the change in PWV is significant. These residuals can be a serious limitation for sub-millimag precision surveys, especially as they are of the same order of amplitude as the transit signals we are looking for. 

In order to differentiate the photometric variations in the differential lightcurves related to changes in PWV from those of astrophysical origin, we implemented a correction as part of the automatic pipeline. First, we needed access to accurate, high cadence PWV measurements, which are provided by LHATPRO. LHATPRO (Low Humidity and Temperature PROfiling radiometer) is a microwave radiometer optimised for measuring PWV (from 0\,mm to a saturation value of 20\,mm, within an accuracy of $\sim$0.1\,mm and with internal precision of 30\,\micro m) situated on a platform at the Very Large Telescope on Cerro Paranal \citep{Kerber2012}. The LHATPRO instrument measures the column of water vapour at zenith approximately every 2 minutes, performs a cone scan at 30$\degree$ for 2.5 minutes every 15 minutes and a 2D all sky scan for 6 minutes every 6 hours. Due to this cone scan there are peaks in the PWV, which we remove, creating small gaps and discontinuities in the PWV measurement. We use a cubic spline to interpolate between the remaining PWV values to get a smooth lightcurve correction. As the gaps are on such a small timescale (of the order of $\sim$5\,min) we don't see it as a concern to the correction.
By using these PWV values, we can then model the effect of the atmospheric absorption with high time resolution on objects of different spectral types (Section \ref{sec51}). This allows us to correct for the differential PWV effect between the target and comparison stars (Section \ref{sec52}).
 
\subsection{Calculating the effect of varying precipitable water vapour on different spectral types}\label{sec51}
To model the effect of the PWV on differential lightcurves, we calculate its `expected' effect on our measurements for objects of different spectral types, observed with the $I+z'$ filter, at different values of PWV and airmass:

\begin{equation} \label{eq:pwv}
f_{I+z'} = \int{W(\lambda,X,V)\, R_{I+z'}(\lambda)\, S(\lambda, T_{\textrm{eff}})~\textrm{d}\lambda}
\end{equation}

\noindent
where $W(\lambda,X,V)$ is the water absorption spectrum at airmass $X$ and precipitable water vapour $V$, $R_{I+z'}$ is the instrument response (including the bandpass for filter $I+z'$, CCD quantum efficiency, CCD window, and reflectivity of the mirror coatings), and $S(\lambda, T_{\textrm{eff}})$ is the synthetic stellar spectrum generated from PHOENIX (\citealt{Husser2013}). This stellar spectrum is dependent on the surface gravity, metallicity, and effective temperature $T_{\textrm{eff}}$ of the star. For simplicity we assumed stars with solar metallicity ($[Fe/H]=0$).

The water absorption spectrum is provided by the SkyCalc Sky Model Calculator, a tool developed by ESO and based on The Cerro Paranal Advanced Sky Model \citep{Jones2013,Noll2012}. This tool provides a library of atmospheric transmission curves for a continuous range of airmass values and discrete PWV values of 0.05, 0.1, 0.25, 0.5, 1.0, 1.5, 2.5, 3.5, 5.0, 10.0, 20.0 and 30.0 mm. We interpolate between these value to create a smooth 4-D grid of all possible values of PWV, airmass,  $T_{\textrm{eff}}$, and $f_{I+z'}$, which can be used to correct any object's differential lightcurve in any frame.

\subsection{Applying PWV correction to differential lightcurves}\label{sec52}

To correct a target differential lightcurve from the effect of PWV, we need to compute its effect on both the target and the artificial reference star. For this purpose, we estimate an effective stellar temperature for the artificial reference star from a weighted mean of the temperatures (extracted from GAIA DR2) of all the comparison stars in the field, using the weights computed by the pipeline in Section \ref{ALC}. The fact that we estimate the temperature of the artificial reference star, and not all of the comparison stars will have a corresponding GAIA DR2 temperature, will have little effect on the correction as most of the calibration stars are G- and K-type. The differential effect between these spectral types is marginal, even for large changes of PWV (see Fig. \ref{fig:pwv_spectype}).

Having a correct estimate of the target's effective temperature is more critical. Inaccuracies in this temperature can lead to over, or under, corrections. Gaia does not provide reliable values for stellar effective temperatures below $3000$\,K \citep{Andrae2018, Dressing2019}, therefore for every target in our target list we carefully estimate its temperature by calculating the absolute $H$-magnitudes for our targets from 2MASS and Gaia and using the $T_{\textrm{eff}}$-magnitude relation in \cite{Filippazzo2015}. These temperature estimates are used as input parameters for the pipeline to compute the effect of the PWV changes on each target's photometry. Finally we divide the PWV effect on the target by the PWV effect on the ALC to generate a differential PWV effect. Then we can correct the target's differential lightcurve by dividing by this differential PWV effect.  

\subsection{Impact and statistics of the PWV correction}
\begin{figure}
 \includegraphics[width=\columnwidth]{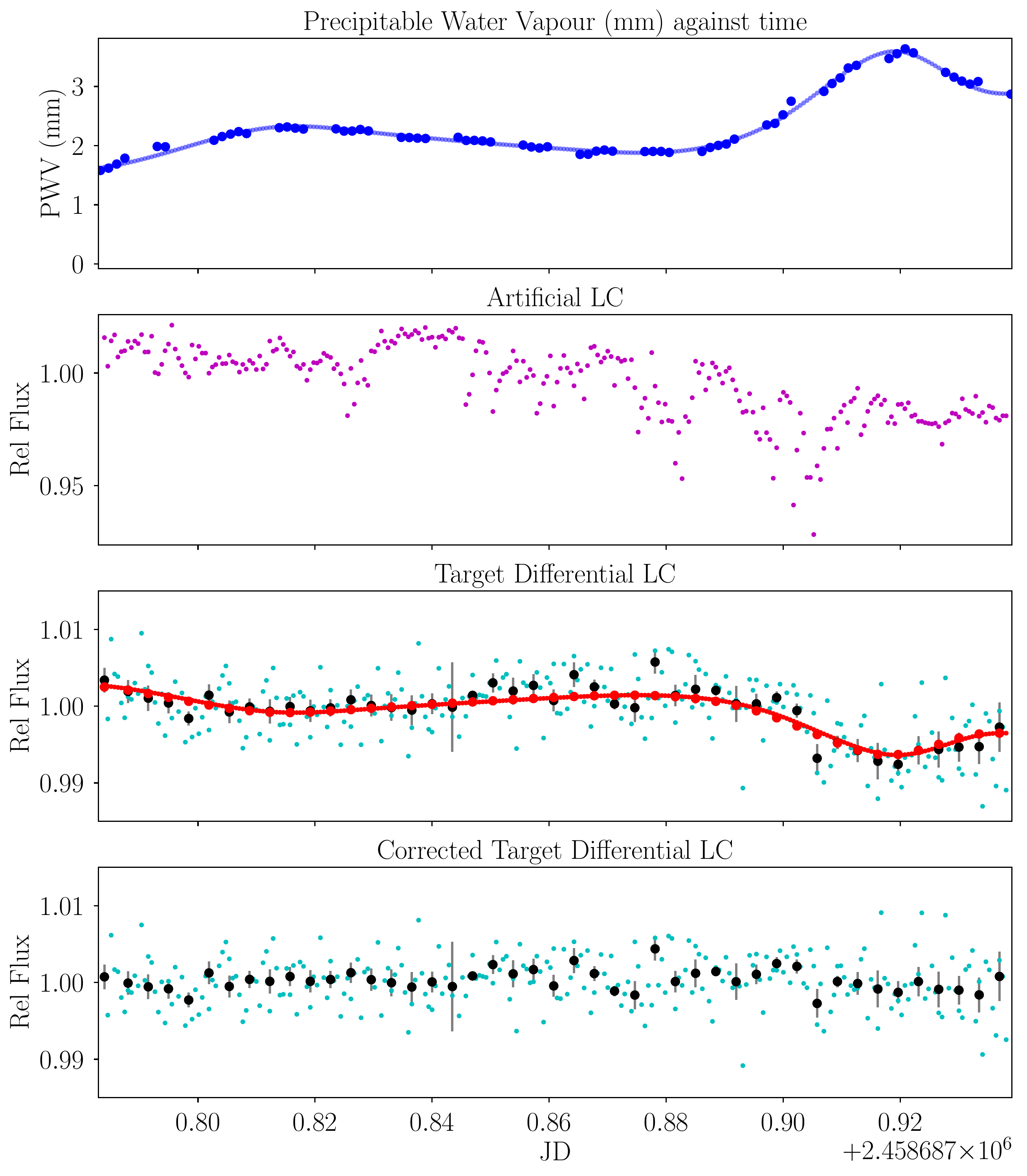}
 \caption{\textbf{Top}: PWV (mm) measurements from LHATPRO for the night of $22^{\textrm{nd}}$ July 2019, with peaks removed. The cubic spline interpolation is shown by the blue line. \textbf{Upper Middle}: The artificial lightcurve generated for this night. \textbf{Lower middle}: Unbinned differential lightcurve (cyan), with 5-min binned points (black), for an M7-type target. A transit-like feature is visible at the end of the lightcurve. The expected differential flux effect of PWV is  shown in red. \textbf{Bottom}: The corrected differential measured lightcurve  in cyan and 5-min binned points in black. We obtain this lightcurve by dividing the original differential lightcurve by the calculated differential flux effect from PWV. The transit-like feature was due to PWV changes and is no longer visible in the corrected lightcurve.}
 \label{fig:faketransit}
\end{figure}

Correction of the PWV effect is a prerequisite to obtain precise differential photometry and to detect shallow transits. This effect impacts the lightcurves over both short (single-night) and long (multi-night) time-scales. During observation of a single night, residuals in the target differential lightcurves may mimic a transit-like signal, even with modest PWV variations of $\sim$1\,mm (see Fig. \ref{fig:faketransit}). 

\begin{figure}
 \includegraphics[width=\columnwidth]{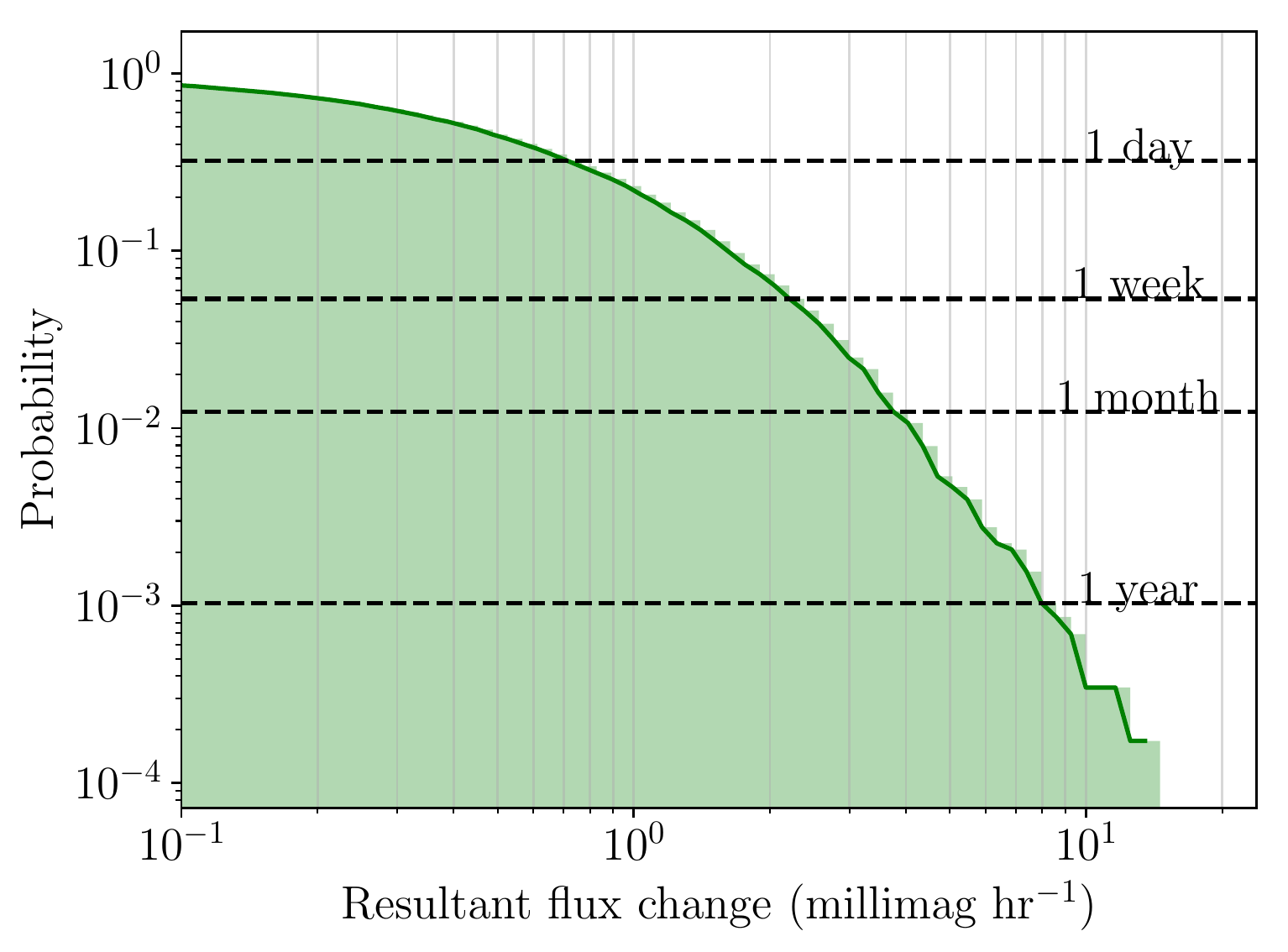}
 \caption{Cumulative histogram of the amplitude change in a target's differential lightcurve induced by PWV variation on typical transit time-scales (1h). We record PWV variations between consecutive 1\,hr bins from 2019 January 1 to 2019 September 18 and used these variations to generate the corresponding differential flux variations for a 2650\,K target object and 4600\,K comparison star. We calculate the amplitude variations that we have a 95 per cent chance of seeing at least one of on a
 daily ($\pm$0.7\,mmag), weekly ($\pm$2\,mmag), monthly ($\pm$4\,mmag) and annual ($\pm$8.1\,mmag) time-scale, marked by the dashed black lines.}
 \label{fig:pwv_hist}
\end{figure}

\begin{figure*}
 \includegraphics[width=\textwidth]{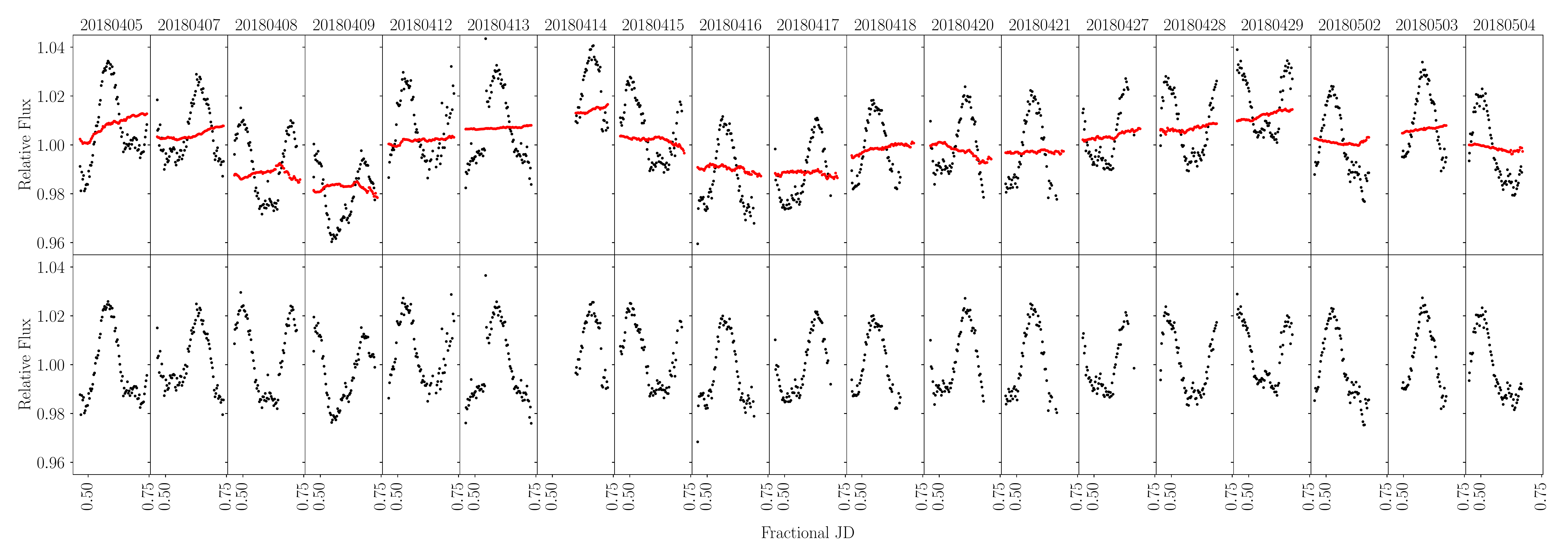}
 \caption{\textbf{Top:} Global $I+z'$ 5-min binned differential lightcurve for an M8-type variable target (LP 609-24, $J$ = 12.33\,mag) is shown in black, observed from 2018 April 5 to 2018 May 6, during the commissioning phase of Callisto. The calculated differential flux effect from PWV is shown in red (5-min binned). This target exhibits both nightly and multi-night variability. \textbf{Bottom:} Water vapour corrected differential lightcurve (5-min binned). While the night-to-night variability remains, the longer time-scale variations were a result of the PWV changes between nights and were removed during the PWV correction.}
 \label{fig:var_correct}
\end{figure*}

By analysing all the PWV measurements from our first year of operation, we can estimate the likelihood of observing a corresponding differential flux effect large enough to be mistaken for a transit. By averaging the PWV values in hour bins (typical time-scale of a transiting planet), the variations between consecutive bins will result in a calculable differential flux effect, for an example 2650\,K target (M7V) and 4600K (K4V) artificial lightcurve. From the cumulative histogram of these differential flux effects (see Fig. \ref{fig:pwv_hist}) we can approximate that we would have a 95 per cent chance of observing at least one flux variation ($\updelta F$) larger than $x$, using: 
\begin{equation}
    P(\updelta F \leq x)^n = 0.05
\end{equation}
where $n$ is the number of flux variations ($n+1$ hour bins) and $P(\updelta F \leq x)$ is the probability of observing a flux variation less than $x$. We estimate that we would have a 95 per cent chance of seeing at least one amplitude variation of $\sim$1\,mmag every night, $\sim$4\,mmag every month, and $\sim$8\,mmag every year. While these larger variations in the lightcurves may not always resemble transits, they are significant enough to affect our detection of a transit, demonstrating the need for our implemented correction.
Over multiple nights of observation, correcting for this effect is an absolute necessity to isolate intrinsic variability of our targets from atmospheric transmission changes due to variation of PWV from one night to another (see Fig. \ref{fig:var_correct}).

\section{Photometric performance of the SSO} \label{performance}

The ability of our automatic pipeline to provide consistent and reproducible results allows us to carry out daily monitoring of the photometric performance and health of the overall system. 

\subsection{Typical photometric precisions of SSO lightcurves}\label{sso_precisions}
\begin{figure}
 \includegraphics[width=\columnwidth]{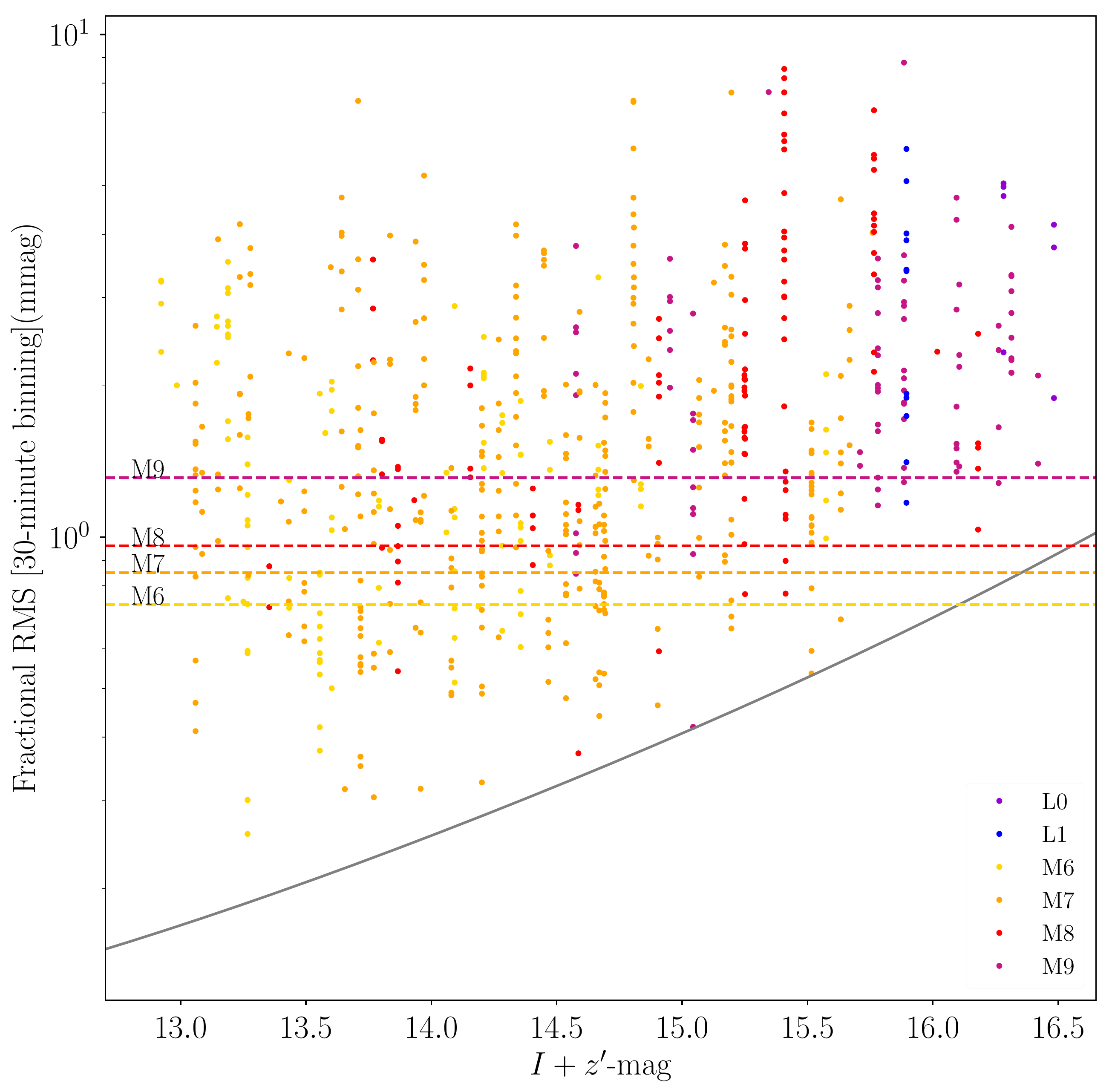}
 \caption{Fractional RMS (for 30-min binning) of all the SSO's UCD target lightcurves carried out with $I+z'$ filter from 2019 January 1 to 2019 September 18. There is a data-point for each target on each night of observation - the vertical lines correspond to different fractional RMS on different nights of observation for the same target. The noise model for the best possible observing conditions is shown in grey. The dashed lines show the minimal level of precision needed to detect a single transit of a TRAPPIST-1b-sized planet (1.127\,$R_{\earth}$) around stars of different spectral types at 9-sigma.}
 \label{fig:rms_izmag}
\end{figure}

To illustrate the typical photometric performances of the facility, and its capability to detect single transits of Earth-size planets, Fig.\,\ref{fig:rms_izmag} displays the measured fractional RMS (for 30-min bins) for the SSO target lightcurves, obtained each night of observation. To ensure there was at least 5 bins for each lightcurve we only included lightcurves where there was more than 150 minutes of total exposure. This accounts for 98 targets and 179 combined nights of observations with multiple telescopes from January 1 to September 18 2019. The binning time-scale we adopted to compute the RMS is set to match the typical transit duration of a short-period planet orbiting an UCD. 

This figure demonstrates that for quiet targets on nights with good observing conditions we are reaching the best possible precision, as determined by our noise model. This noise model \citep{Merline1995} accounts for several different contributions: Poisson noise from the star, read noise from the detector, noise from background light, noise from dark current, and atmospheric scintillation. For the atmospheric scintillation, we use a modified form of Young's approximation, specific for Paranal \citep{Osborn2015}. The targets we observe typically have exposure times from 10--60\,s, therefore we assume the noise model for 60\,s exposure, with an overhead of 10.5\,s, which gives 25 data-points in each 30-min bin. The noise model illustrated in Fig. \ref{fig:rms_izmag} is also assumed for an aperture of 11.3 pixels on the best possible night, with an airmass of 1 and a background sky level of 49.04 ADU\,pixel$^{-1}$ (the lowest recorded sky background since January 2019). 

There is no correction for photometric variability, removal of bad weather, or selection of the nights with the best observing conditions. This results in the vertical stripes for each target corresponding to large spreads in RMS in the lightcurves for different nights, related to the wide range of observing conditions and potentially that target's intrinsic variability. This spread can be seen to limit our single transit detection efficiency, thereby demonstrating the need to remove photometric variability. We expect the median precision we find (and our detection potential) to improve when the stellar variability is properly accounted for, which will be presented in a future paper.

The photometric precisions reached by our least active targets in this diagram show that we are reaching sub-millimag precisions for approximately 30 per cent of lightcurves (with a median precision of $\sim$1.5\,mmag), and up to $\sim$0.26\,mmag for the brightest objects. In Fig.\,\ref{fig:rms_izmag}, we superimposed an approximation of the minimum photometric precision required to measure a single transit by a TRAPPIST-1b size planet (1.127\,$R_{\earth}$) with a signal-to-noise ratio of 9 for different spectral types \citep{Pecaut2013}. This demonstrates SSO's excellent quality and detection capability, especially for quiet targets observed on nights with good observing conditions. 

\subsection{Simultaneous observation comparison with TESS}
NASA's Transiting Exoplanet Satellite Survey (TESS, \citealt{Ricker2015}) was launched in April 2018. While TESS is optimised for detecting planets around G to mid M-dwarf stars, their wide bandpass allows them to additionally observe the brightest late M-dwarfs with high precisions.

\begin{figure}
 \includegraphics[width=\columnwidth]{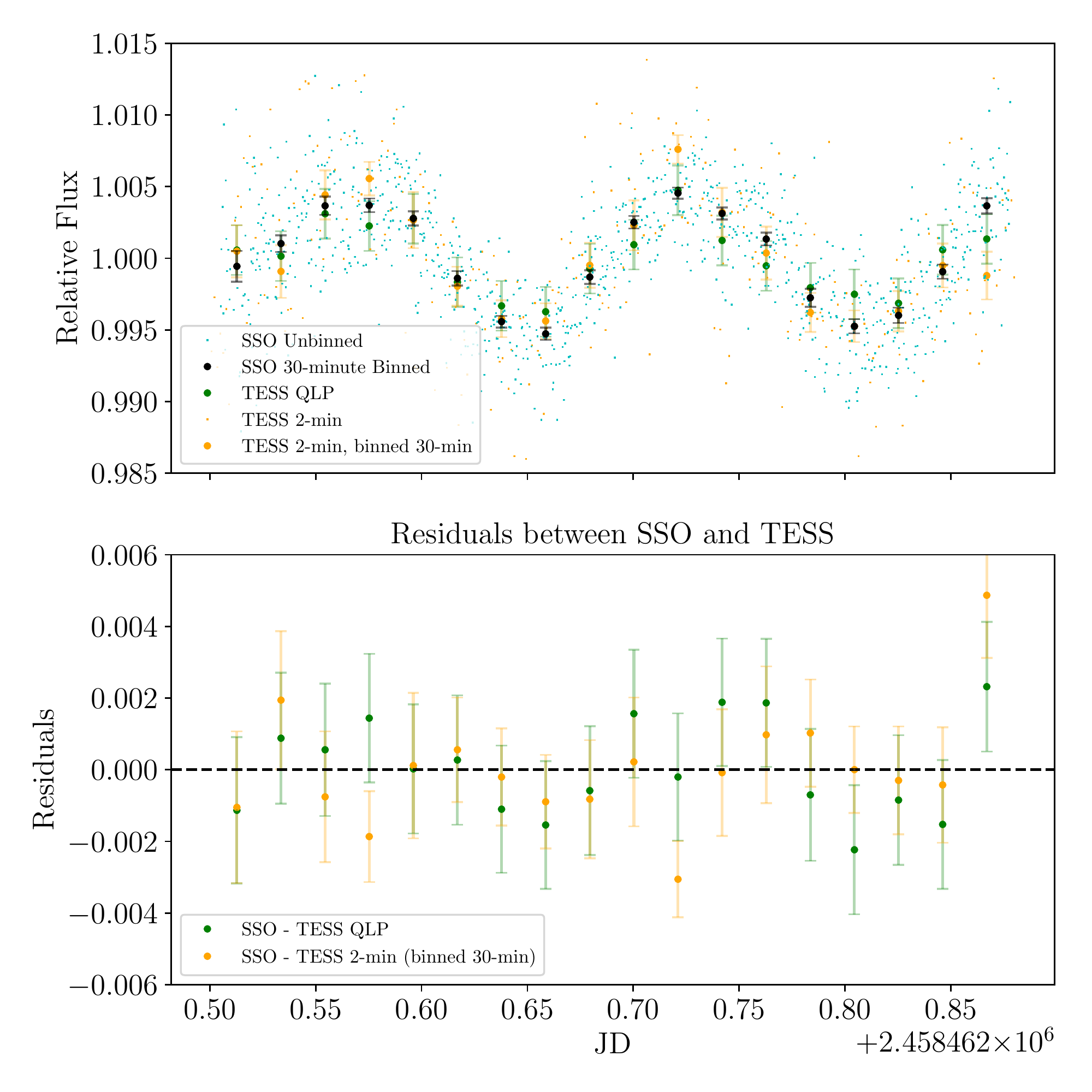}
 \caption{\textbf{Top}: SSO's differential lightcurve compared to the lightcurves from TESS 2-minute cadence data and MIT QLP 30-minute cadence data for an M6V object ($J=10.3$\,mag) on 2018 December 10. \textbf{Bottom}: The residuals between the TESS and SSO lightcurves.}
 \label{fig:TESS}
\end{figure}

Here we present a comparison of a night of simultaneous observation of the M6 star, WOH G 618 (TIC 31381302, $J=10.3$\,mag, $T = 12.5$\,mag, $I+z' = 12.6$\,mag), by a single SSO telescope to TESS data (Fig. \ref{fig:TESS}). For TESS, we include both the publicly available 2-minute cadence data and the final lightcurve from the MIT Quick Look Pipeline (QLP). The QLP was developed to extract lightcurves specifically for targets in the 30-minute full frame images (FFIs). It is shown here as an example for FFI photometry, allowing us to gauge the precision that can be achieved for targets which are not part of TESS' 2-minute sample. The QLP and other custom pipelines can be used to extract lightcurves from the FFIs for the majority of late M-dwarfs in the TESS fields.

We see excellent agreement between the three data-sets. There remains structure in the SSO-TESS QLP residuals that appears to correlate with the variability, however, this is within error. The SSO lightcurve shows less white noise than TESS, as expected because TESS is not optimised for these very red objects. For fainter and redder UCDs we expect that the quality of the SSO lightcurves will exceed TESS, however, for the brightest SPECULOOS targets the lightcurves will be comparable. We believe this demonstrates the remarkable performance of both TESS and SSO, especially considering the detection potential when combining simultaneous observations from multiple SSO telescopes and TESS together.

\section{Discussion and Perspectives}

This paper illustrates a practical and successful implementation of an automated differential photometry algorithm with carefully calibrated weighting schemes for comparison stars, and a correction of the effect of varying telluric water vapour. The analysis of the photometric performance of SSO's first year of operation shows that, with these methods, we can regularly reach sub-millimag precision photometry, for our quieter targets.

Several publications have already addressed this telluric water vapour problem when observing cool stars in the near-IR \citep{Bailer-Jones2003, Blake2008, Blake2011}. The MEarth survey has a similar 715--1000\,nm bandpass and also witnessed induced photometric systematics that could mimic an exoplanet transit, due to variations in atmospheric water vapour \citep{Berta2012}. These systematics were also a limiting factor in the type of M-dwarfs ($>0.15\,R_{\sun}$) that could have rotation periods extracted from MEarth \citep{Newton2018}. 

Despite identification of the issue, we have not found any implemented correction of telluric water vapour, directly from first principles, for a large-scale survey in the literature. However, we do note that MEarth developed an alternative method of correcting the water vapour effect \citep{Irwin2011}, by medianing all M dwarf lightcurves gathered by their 8 telescopes (at each site) in half hour time bins to create a ``common mode'' lightcurve. They then calculate a scaling factor for each star, determined by a least-squares optimisation. While this method has proved successful to a survey like MEarth \citep{Berta2011}, which observes dozens of stars every 30 minutes, for SSO, which only ever observes a maximum of 4 M dwarf targets at once, this technique is limited. We believe that correcting the water vapour from the transmission spectra directly offers the advantage that it is determined from an independent dataset (LHATPRO), and removes the chance of overfitting real structure.

It was therefore necessary to develop a model to correct for this differential effect. Additionally, this work highlights how beneficial it is to have access to high time resolution, high precision PWV measurements. This correction, however, has a wider impact than just the correction of the SSO lightcurves. It could be applied to any future transit survey observing redder stars in the near-IR, including earlier M dwarfs, or more generally, for example in long-term photometric variability studies of red objects.

Not every facility has access to expensive water vapour radiometers and so there has been substantial development of alternate methods of measuring the PWV. Instruments like aTmCam \citep{Li2012,Li2014} and CAMAL \citep{Baker2017} use a set of imagers to take simultaneous observations of bright calibration stars with different narrow-band filters chosen to be in-band and out-of-band for water.
Along with measurements of local surface pressure and temperature, GPS receivers have also been used to estimate the atmospheric PWV to accuracies of 0.11--1\,mm \citep{Bevis1992,Duan1996,Blake2011,Castro2016,Li2018}. We have shown in this paper that changes in PWV of 1\,mm are sufficient to limit our detection efficiency and can even mimic a transit from an Earth-sized planet, so accurate PWV measurements are essential. As an alternative to correcting the effect, it is possible to minimize the impact of water bands in the near-IR (and the photometric consequences from changing PWV) by reducing the filter band-pass, but at the cost of losing stellar photons and the need for a larger telescope. 

We have identified a couple of limitations in our PWV correction, that could potentially leave some residual structures in the final differential lightcurve. The LHATPRO instrument saturates at 20\,mm at zenith which will limit the accuracy we can achieve for very high PWV, especially for high airmass. There is also a $\sim$200\,m vertical distance between the VLT platform (2635m) and the SSO facility. Additionally, the LHATPRO instrument measures the water vapour at zenith instead of along our line-of-sight. All of these factors may result in underestimating the amount of PWV affecting our observations. The effect on our photometry is, however, likely to be small; \cite{Querel2014} found that PWV over Paranal was spatially homogeneous down to elevations of 27.5\degree, such that measuring PWV along zenith is sufficient for most astronomical applications. Concerningly, this homogeneity was found to decrease with rising levels of water in the atmosphere, as they found the PWV variations were reliably 10--15 per cent of the absolute PWV. Therefore our correction is likely to be most effective at zenith where we don't have to consider spatial variations, and more effective at low values of PWV (<2\,mm), where the variations across the sky are of the order $\sim$0.1--0.3\,mm. An investigation of the impact from these various effects on our precise photometry is planned for the future.

As mentioned in Section \ref{sso_precisions}, stellar variability can seriously limit our planet detection efficiency. Future development of the pipeline will essentially focus on the implementation of an algorithm to identify and model flares and variability simultaneously with an automatic transit search. By optimising our detection efficiency SPECULOOS provides a unique opportunity to explore the planetary population around UCDs, matching space-level photometric precisions with an ability to study fainter and redder objects than ever before.
\section*{Acknowledgements}
The research leading to these results has received funding from the European Research Council (ERC) under the FP/2007--2013 ERC grant agreement n$^{\circ}$ 336480, and under the H2020 ERC grants agreements n$^{\circ}$ 679030 \& 803193; and from an Actions de Recherche Concert\'{e}e (ARC) grant, financed by the Wallonia--Brussels Federation. This work was also partially supported by a grant from the Simons Foundation (PI Queloz, grant number 327127), as well as by the MERAC foundation (PI Triaud). LD acknowledges support from the Gruber Foundation Fellowship. MG and EJ are F.R.S.-FNRS Senior Research Associates. B.-O.D. acknowledges support from the Swiss National Science Foundation (PP00P2-163967). VVG is a F.R.S.-FNRS Research Associate. MNG and CXH acknowledge support from Juan Carlos Torres Fellowships.




\bibliographystyle{mnras}
\bibliography{ref}



\bsp	
\label{lastpage}
\end{document}